# Non-negative Rational Semantic Numeration Systems

**Alexander Yu. Chunikhin**
**Palladin Institute of Biochemistry**
**National Academy of Sciences of Ukraine**
alexchunikhin61@gmail.com
ORCID 0000-0001-8935-0338

**Abstract**. A new class of Semantic Numeration Systems, namely, positive rational **SNS(Q⁺)** is introduced. For cardinal semantic operators of the form (↑**Q⁺**), differences in the formation of a carry (common carry) and remainders are defined. The properties of **SNS(Q⁺)** as dynamic systems are formulated and illustrated through analytical and numerical examples. A first attempt at defining partial integer Semantic Numeration Systems in the form of **SNS(q⁻)** is proposed.

## 1. Introduction

Recently, there has been a growing interest in rational numeration systems, i.e., systems with a base $\frac{p}{q}$, where $p$ and $q$ are coprime integers ($p > q$). Considered the seminal work in this area, (Akiyama, Frougny, & Sakarovitch, 2008) establishes that every positive integer has a unique expansion in a rational numeration system, and every real number has at least one expansion. One of the latest works (Andrieu, Eliahou, & Vivion, 2025) is devoted to studying families of infinite words, called minimal and maximal words, and to examining their richness threshold and discrepancy.

It is necessary to emphasize at least two aspects that distinguish traditional numeration systems from semantic ones.

(i) In most traditional works, the representation of numbers in a certain numeration system is treated as having been accomplished, completed. In the author's works, the representation of a number (more precisely, a multinumber) is presented as a sequence of transformations – a dynamic process in a certain state space.

(ii) Traditional numeration systems consider only the linear topology of representation (expansion). In Semantic Numeration Systems various topologies are allowed, with possible heterogeneity and anisotropy.

This paper proposes an extension of Semantic Numeration Systems (Chunikhin, 2021; 2022; 2025) for the case in which not only the radix (base) but also all components of the cardinal semantic transformation are non-negative rational numbers.

## 2. The theoretical foundation of SNS

This section first presents the basic concepts of the semantic numeration systems theory (Chunikhin, 2021; 2022; 2025) and then introduces the specifics of non-negative rational numeration.

An *abstract entity* (Æ) is an entity of arbitrary nature provided with an identifier name that allows it to be distinguished from other entities. *Cardinal Abstract Entity* (CÆ) is an abstract entity with a cardinal characteristic $CÆ_i = (i; \#_i)$, where $i$ is the name of the cardinal abstract entity, $\#_i = \text{Card}(CÆ_i)$, $\#_i \in \mathbf{N_0}$.

*Multeity* is the manifestation of something essentially uniform in various kinds and forms as well as the quality or condition of being multiple or consisting of many parts. Since we will further deal with the transformation of meanings, we define the corresponding specific type of multeity as semantic. *Semantic multeity* is an abstract space with no more than a countable set of abstract entities, semantically united by a context. *Cardinal Semantic Multeity* (CSM) is a semantic multeity, each element of which is equipped with a cardinal characteristic – the multiplicity of a given abstract entity represented in multeity. From a set-theoretic point of view, a cardinal semantic multeity is a multiset, the carrier of which is contextually conditioned. The elements of the cardinal semantic multeity are *cardinal abstract entities*.

*Cardinal Semantic Operator* (CSO) is a multivalued mapping of the cardinal semantic multeity on itself, which associates a set of entity-operands from the multeity with a set of entity-images from the same multeity, transforming their cardinals using the operations defined by the operator signature: $\text{Signt}(CSO) = (K, \text{Form}, |n\rangle_w, |r\rangle_v)$, where $K$ is the operator kind, *Form* is the operator type, $|n\rangle$ is a radix-vector, $|r\rangle$ is a conversion vector. The pair (W, V) is a valence of the cardinal semantic operator.

The kind of the cardinal semantic operator indicates the content of the transformations (definition of a carry and a remainder), which are performed with the cardinals of CÆ-operands. The generally accepted form of the carry and remainder formation in traditional (radix-multiplicity, designated here as (↑#)) numeration systems is: the carry $p_i = \lfloor \#_i / n_i \rfloor$, the remainder: $\text{rem } \#_i = \#_i - p_i \cdot n_i$. The theoretical foundations of SNS, formed by cardinal semantic operators of the form (↑#), in which all parameters and quantities take non-negative integer values, are given in (Chunikhin, 2021).

In this paper, we propose an extension of SNS to "rational plus", that is, formed by cardinal semantic operators of the form (↑$\mathbf{Q^+}$), in which all parameters and quantities take non-negative rational values. In such numeration systems, we assume that the carry is formed without using the floor function $\lfloor . \rfloor$: $p_i = \#_i / n_i$; and the remainder: $\text{rem } \#_i = \#_i - p_i \cdot n_i$, ($p_i$, $\text{rem } \#_i \in \mathbf{Q^+}$).

The main *forms* of the cardinal semantic operators $CSO(\mathbf{Q^+})$ are:

➢ *L-operator*($\mathbf{Q^+}$) (Linear operator): (↑$\mathbf{Q^+}$, L, $n_i$, $r_{ij}$) is a cardinal semantic operator of valency (W, V) = (1, 1), which determines the carry value $p_i$ (radix-$n_i$ $\mathbf{Q^+}$-multiplicity) from the cardinal abstract entity $CÆ_i$ and assigns (gives meaning to) the value $r_{ij} \cdot p_i$ to the transformant $q_j$ added to the cardinal $\#_j$ of the abstract entity $CÆ_j$.

A diagrammatic representation of the L-operator($\mathbf{Q^+}$) is shown in Fig. 1.

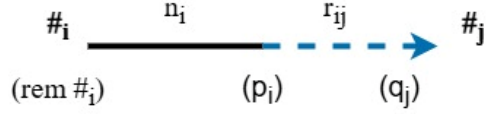

Fig. 1

When the L-operator($Q^+$) acts on the $CÆ_i$-operand, the following operations are performed:
(i)   $p_i = \#_i/n_i$ – calculation of the radix-multiplicity, that is, *i-carry* value;
(ii)  rem $\#_i = \#_i - p_i \cdot n_i = 0$ – calculation of the *remainder* in $CÆ_i$;
(iii) $q_j = r_{ij} \cdot p_i$ – calculation of the *j-transformant* value;
(iv)  $\#_j` = \#_j + q_j$ – calculation of the resulting cardinal in the $CÆ_j$-image.

➢ *D-operator*($Q^+$) (Distribution operator): ($\uparrow Q^+$, D, $n_i$, ($r_{ij}$, …, $r_{ih}$)) is a cardinal semantic operator of valency (W, V) = (1, v), which determines the carry value $p_i$ from the cardinal abstract entity $CÆ_i$ and assigns the values to v transformants as follows: the value $r_{ij} \cdot p_i$ to the transformant $q_j$ added to the cardinal $\#_j$ of the abstract entity $CÆ_j$, …, the value $r_{ih} \cdot p_i$ to the transformant $q_h$ added to the cardinal $\#_h$ of the abstract entity $CÆ_h$.

A diagrammatic representation of the D-operator($Q^+$), namely, $D_2$ is shown in Fig. 2.

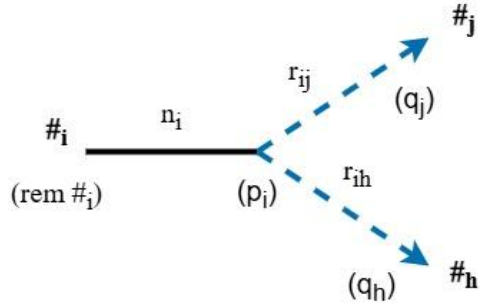

Fig. 2

When the $D_2$-operator($Q^+$) acts on the $CÆ_i$-operand, the following operations are performed:
(i)   $p_i = \#_i/n_i$ – calculation of the radix-multiplicity, that is, *i-carry* value;
(ii)  rem $\#_i = \#_i - p_i \cdot n_i = 0$ – calculation of the *remainder* in $CÆ_i$;
(iii) $q_j = r_{ij} \cdot p_i$, $q_h = r_{ih} \cdot p_i$ – calculation of the *j, h-transformants* value;
(iv)  $\#_j` = \#_j + q_j$, $\#_h` = \#_j + q_h$ – calculation of the resulting cardinals in the $CÆ_j$-, $CÆ_h$-images.

➢ *F-operator*($Q^+$) (Fusion operator): ($\uparrow Q^+$, F, ($n_i$, …, $n_j$), $r_{(i…j)h}$) is a cardinal semantic operator of valency (W, V) = (w, 1), which, for *common carry* value $p_{(i…j)}$ from the abstract entities $CÆ_i$, …, $CÆ_j$, assigns the value $r_{(i…j)h} \cdot p_{(i…j)}$ to the transformant $q_h$ added to the cardinal $\#_h$ of the abstract entity $CÆ_h$.

A diagrammatic representation of the F-operator($Q^+$), namely, $_2F$ is shown in Fig. 3.

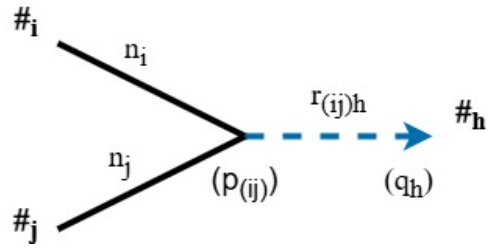

Fig. 3

When the $_2$F-operator($Q^+$) acts on the $C\!Æ_i$-, $C\!Æ_j$-operands, the following operations are performed:
(i)   $p_i = \#_i/n_i$, $p_j = \#_j/n_j$ – calculation of the *partial carries*;
(ii)  $p_{(ij)} = \min(p_i, p_j)$ – calculation of the *common carry*;
(iii) rem $\#_i = \#_i - p_{(ij)} \cdot n_i$, rem $\#_j = \#_j - p_{(ij)} \cdot n_j$ – calculation of the *remainders* in $C\!Æ_i$ and $C\!Æ_j$;
(iv)  $q_h = r_{(ij)h} \cdot p_{(ij)}$ – calculation of the *h-transformant* value;
(v)   $\#_h` = \#_h + q_h$ – calculation of the resulting cardinal in the $C\!Æ_h$-image.

From (iii) it follows that for $C\!Æ$, whose partial carry is chosen as the common one, the remainder is always zero: $p_{(ij)} = \min(p_i, p_j) = p_i \Rightarrow$ rem $\#_i = 0$.

**Example 1.** Let $\#_i = 7$, $\#_j = 3$, $n_i = 1/3$, $n_j = 2/5$, $r_{(ij)h} = 2/7$, $\#_h = 1$.
Then, $p_i = 21$, $p_j = 15/2 \Rightarrow p_{(ij)} = \min(21, 15/2) = 15/2$;
$$\text{rem } \#_i = 7 - (1/3) \cdot (15/2) = 9/2 = 4.5, \text{ rem } \#_j = 0;$$
$q_h = (2/7) \cdot (15/2) = 15/7$, $\#_h` = 1 + 15/7 = 22/7$.

➤ *M-operator*($Q^+$) (Multi-operator): ($\uparrow Q^+$, M, $(n_i, \ldots, n_j)$, $(r_{(i\ldots j)h}, \ldots, r_{(i\ldots j)g})$) is a cardinal semantic operator of valency (W, V) = (w, v), which for *common carry* value $p_{(i\ldots j)}$ from the $w$ abstract entities $C\!Æ_i, \ldots, C\!Æ_j$, assigns the values to $v$ transformants as follows: the value $r_{(i\ldots j)h} \cdot p_{(i\ldots j)}$ to the transformant $q_h$ added to the cardinal $\#_h$ of the abstract entity $C\!Æ_h$, …, the value $r_{(i\ldots j)g} \cdot p_{(i\ldots j)}$ to the transformant $q_g$ added to the cardinal $\#_g$ of the abstract entity $C\!Æ_g$.

A diagrammatic representation of the M-operator($Q^+$), namely, $_2M_2$ is shown in Fig. 4.

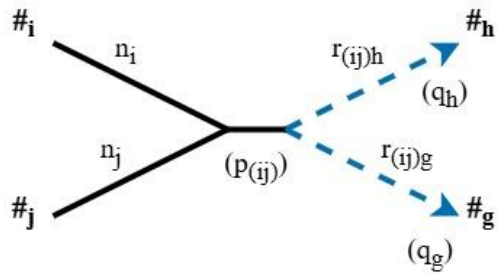

Fig. 4

When the $_2M_2$-operator($Q^+$) acts on $CÆ_i$-, $CÆ_j$-operands, the following operations are performed:

(i) $p_i = \#_i/n_i$, $p_j = \#_j/n_j$ – calculation of the *partial carries*;
(ii) $p_{(ij)} = \min(p_i, p_j)$ – calculation of the *common carry*;
(iii) $\text{rem } \#_i = \#_i - p_{(ij)} \cdot n_i$, $\text{rem } \#_j = \#_j - p_{(ij)} \cdot n_j$ – calculation of the *remainders* in $CÆ_i$ and $CÆ_j$;
(iv) $q_h = r_{(ij)h} \cdot p_{(ij)}$, $q_g = r_{(ij)g} \cdot p_{(ij)}$ – calculation of the *partial transformants*;
(v) $\#_h` = \#_h + q_h$, $\#_g` = \#_g + q_g$ – calculation of the resulting cardinals in the $CÆ_h$-, $CÆ_g$-image.

From (iii) it follows that for $CÆ$, whose partial carry is chosen as the common one, the remainder is always zero: $p_{(ij)} = \min(p_i, p_j) = p_i \Rightarrow \text{rem } \#_i = 0$.

These four cardinal semantic operators form the operator basis of any semantic numeration system.

To represent complex multistage semantic transformations, mono-operator transformations are typically insufficient. In (Chunikhin, 2021) the concept of a *Numeration Space* (NS) was introduced, the elements of which are Cardinal Abstract Objects (CAO). *A Cardinal Abstract Object* (CAO) is a set of Cardinal Abstract Entities connected in a certain topology (STop) by Cardinal Semantic Operators. A certain $CAO_I$ implements/represents a specific method of numeration *I* in a numeration space. *I* is a name (may be a complex name) denoting the $CAO_I$ as a specific *numeration method* in accordance with the accepted classification (Chunikhin, 2021).

*Cardinal Semantic Transformation* (CST) consists in executing all "active" cardinal semantic operators for the given $CAO_I$. A *CST step* means a single execution of all "active" cardinal semantic operators for the given $CAO_I$.

A holistic structural-cardinal representation of the $CAO_I$ after the *k*-th step of the cardinal semantic transformation is called a *I-multinumber* of the step *k* and denoted by $A_I(k)$.

$$A_I(k) = \mathbb{L}^m(|\#(k)\rangle \mid Ᵽ_I), \qquad (1)$$

where
- $\mathbb{L}^m$ is the symbol of the structural-cardinal union in the $CAO_I$ of both *m* cardinal abstract entities and the cardinal semantic operators that connect them;
- $|\#(k)\rangle$ is the *CÆ-vector* formed by *m* components of the multicardinal $<A_I(k)>$;
- $Ᵽ_I$ is the $CAO_I$ *configuration matrix* ($m \times m$), containing information about the type of the cardinal semantic operators in the $CAO_I$, their parameters and connectivity topology (STop).

A *Semantic Numeration System* is defined as a *class of numeration methods* that are homogeneous in terms of certain classification features (Chunikhin, 2021).

### 3. Semantic Numeration Systems ($Q^+$) as Dynamical Systems

Let us consider an arbitrary class of cardinal abstract objects **CAO** as a set of structurally homogeneous dynamic systems, differing only in dimensionality and parameter values. This allows us to present the central principles of representing semantic numeration systems as dynamical systems, using an arbitrary $CAO_I$ from the given class as an example.

We define the subspace $CSM_I \subseteq \mathbf{CSM}$ as the $m$-dimensional state space of the given $CAO_I$. Each $i$-coordinate of the space is formed by a state variable, namely, a cardinal abstract entity $CÆ_i \in CSM_I$. The multicardinal $\langle\#(k)\rangle$ in this formulation acquires the meaning of the *state vector* of the $CAO_I$: $|\#(k)\rangle$, $\dim(|\#(k)\rangle) = m$.

At each step of the CST, the cardinal of any $CÆ_i \in CAO_I$ is determined by its current value minus what goes out as a carry, plus the transformants from the incoming CSOs. Then, for the $CAO_I$ as a whole, the state equation takes the following form:

$$|\#(k+1)\rangle = |\#(k)\rangle - \mathbf{N}|p.(k)\rangle + \mathbf{R}^T|p.(k)\rangle,$$

where the common carry vector $|p.(k)\rangle$ (the control vector) for $CAO(\mathbf{Q}^+)$ is defined as

$$|p.(k)\rangle = \Lambda \mathbf{N}^-|\#(k)\rangle.$$

Here $\Lambda$ is a *common carry operator*:

$$|p.(k)\rangle = \Lambda |p(k)\rangle.$$

The *radix operator* $\mathbf{N}$ is a diagonal matrix ($m \times m$) with elements $n_i \delta_{ij}$. The *inverse radix operator* $\mathbf{N}^-$ is a diagonal matrix ($m \times m$) with elements $n_i^{-1} \delta_{ij}$. The *conversion operator* $\mathbf{R}^T$ is a matrix ($m \times m$) that contains the conversion coefficients transposed relative to their initial positions in the configuration matrix. All four of the above operators are formed based on the configuration matrix Ᵽ (1).

Then, we have the following state equation of the $CAO(\mathbf{Q}^+)$:

$$|\#(k+1)\rangle = |\#(k)\rangle + (\mathbf{R}^T - \mathbf{N})\Lambda \mathbf{N}^-|\#(k)\rangle. \qquad (2)$$

In cases where the $CAO(\mathbf{Q}^+)$ contains only $L(\mathbf{Q}^+)$- and $D(\mathbf{Q}^+)$-cardinal semantic operators, the common carry operator $\Lambda$ is not needed in the formation of the carry vector: $|p.(k)\rangle = |p(k)\rangle$. Then,

$$|\#(k+1)\rangle = |\#(k)\rangle + (\mathbf{R}^T - \mathbf{N})\mathbf{N}^-|\#(k)\rangle. \qquad (3)$$

The equation (2) is universally applicable, meaning it can be used to describe the dynamics of the $CAO(\mathbf{Q}^+)$ with an arbitrary set of cardinal semantic operators, their parameters and connectivity topology.

In the case of a stationary CAO the parameters of the CAO don't change, and we have the state equation of the $CAO(\mathbf{Q}^+)$ in the form (2). In cases where a stationary $CAO(\mathbf{Q}^+)$ contains only L- and D-cardinal semantic operators, we have the state equation of the $CAO(\mathbf{Q}^+)$ in the form (3).

In the case of a non-stationary CAO the parameters of the CAO change, in general, at each step of the CST ($R^T \rightarrow R^T(k), N \rightarrow N(k), N^- \rightarrow N^-(k)$). The state equation for a non-stationary CAO($Q^+$) takes the form:

$$|\#(k+1)\rangle = |\#(k)\rangle + (R^T(k) - N(k))\Lambda N^-(k)|\#(k)\rangle. \tag{4}$$

## 4. Example 2

Let us examine in more detail the components of the state equation (2) of some CAO$_I$($Q^+$) using a specific example (Fig.5). We denote the initial CÆs by triangles with upward-pointing vertices, the intermediate ones by circles, and the final one by a downward-pointing triangle. This CAO$_I$($Q^+$) is composed of seven CÆs and four CSOs – $_2$M$_2$, L, D$_2$ and $_2$F – that connect them in the given STop. For generality, all types of cardinal semantic operators are used.

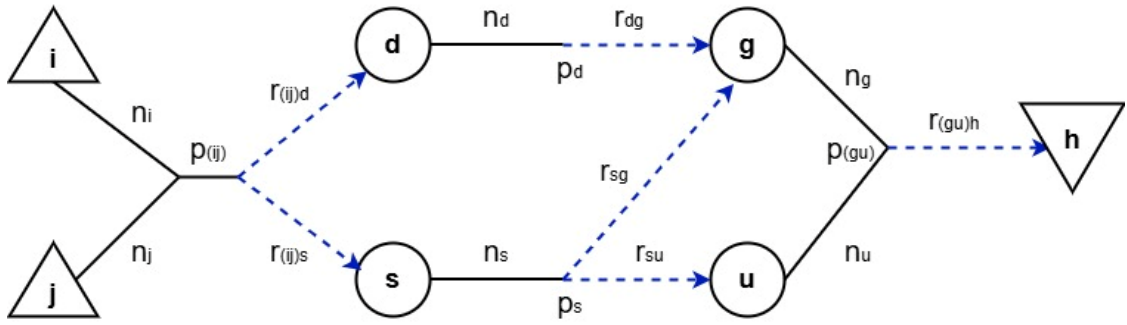

Fig. 5

The state vector (the multicardinal in vector form) for the given CAO$_I$($Q^+$) has the following form ($m = 7$):

$$|\#(k)\rangle = \begin{pmatrix} \#_i(k) \\ \#_j(k) \\ \#_d(k) \\ \#_s(k) \\ \#_g(k) \\ \#_u(k) \\ \#_h(k) \end{pmatrix}.$$

The configuration matrix for the considered CAO$_I$ is as follows:

$$\mathbb{R} = \begin{array}{c|ccccccc} \backslash & i & j & d & s & g & u & h \\ \hline i & n_i & 0 & r_{(ij)d} & r_{(ij)s} & 0 & 0 & 0 \\ j & 0 & n_j & r_{(ij)d} & r_{(ij)s} & 0 & 0 & 0 \\ d & 0 & 0 & n_d & 0 & r_{dg} & 0 & 0 \\ s & 0 & 0 & 0 & n_s & r_{sg} & r_{su} & 0 \\ g & 0 & 0 & 0 & 0 & n_g & 0 & r_{(gu)h} \\ u & 0 & 0 & 0 & 0 & 0 & n_u & r_{(gu)h} \\ h & 0 & 0 & 0 & 0 & 0 & 0 & 0 \end{array}.$$

From the operators $R^T$ and $N$ we immediately express the difference:

$$R^T - N = \begin{array}{c|ccccccc} \backslash & i & j & d & s & g & u & h \\ \hline i & -n_i & 0 & 0 & 0 & 0 & 0 & 0 \\ j & 0 & -n_j & 0 & 0 & 0 & 0 & 0 \\ d & r_{(ij)d} & 0 & -n_d & 0 & 0 & 0 & 0 \\ s & 0 & r_{(ij)s} & 0 & -n_s & 0 & 0 & 0 \\ g & 0 & 0 & r_{dg} & r_{sg} & -n_g & 0 & 0 \\ u & 0 & 0 & 0 & r_{su} & 0 & -n_u & 0 \\ h & 0 & 0 & 0 & 0 & r_{(gu)h} & 0 & 0 \end{array}.$$

The inverse radix operator can be written as

$$N^- = \begin{array}{c|ccccccc} \backslash & i & j & d & s & g & u & h \\ \hline i & n_i^{-1} & 0 & 0 & 0 & 0 & 0 & 0 \\ j & 0 & n_j^{-1} & 0 & 0 & 0 & 0 & 0 \\ d & 0 & 0 & n_d^{-1} & 0 & 0 & 0 & 0 \\ s & 0 & 0 & 0 & n_s^{-1} & 0 & 0 & 0 \\ g & 0 & 0 & 0 & 0 & n_g^{-1} & 0 & 0 \\ u & 0 & 0 & 0 & 0 & 0 & n_u^{-1} & 0 \\ h & 0 & 0 & 0 & 0 & 0 & 0 & 0 \end{array}.$$

The common carry operator is

$$\Lambda = \begin{array}{c|ccccccc} \backslash & i & j & d & s & g & u & h \\ \hline i & \Lambda & \Lambda & 0 & 0 & 0 & 0 & 0 \\ j & \Lambda & \Lambda & 0 & 0 & 0 & 0 & 0 \\ d & 0 & 0 & 0 & 0 & 0 & 0 & 0 \\ s & 0 & 0 & 0 & 0 & 0 & 0 & 0 \\ g & 0 & 0 & 0 & 0 & \Lambda & \Lambda & 0 \\ u & 0 & 0 & 0 & 0 & \Lambda & \Lambda & 0 \\ h & 0 & 0 & 0 & 0 & 0 & 0 & 0 \end{array}.$$

The partial carries vector $|p(k)\rangle$ is

$$|p(k)\rangle = \begin{pmatrix} p_i(k) \\ p_j(k) \\ p_d(k) \\ p_s(k) \\ p_g(k) \\ p_u(k) \\ 0 \end{pmatrix}.$$

The action of the operator that forms the common carry consists in "multiplying" the matrix $\Lambda$ by the vector of partial carries $|p(k)\rangle$ in such a way that the operation of adding the factors is replaced by the operation of taking their minimum.

Then the vector of common carries $|p.(k)\rangle$ (the control vector) is defined as (the step $k$ is omitted here):

$$\begin{pmatrix} p_{(ij)} \\ p_{(ij)} \\ p_d \\ p_s \\ p_{(gu)} \\ p_{(gu)} \\ 0 \end{pmatrix} = \begin{array}{c|ccccccc} \backslash & i & j & d & s & g & u & h \\ i & \Lambda & \Lambda & 0 & 0 & 0 & 0 & 0 \\ j & \Lambda & \Lambda & 0 & 0 & 0 & 0 & 0 \\ d & 0 & 0 & 0 & 0 & 0 & 0 & 0 \\ s & 0 & 0 & 0 & 0 & 0 & 0 & 0 \\ g & 0 & 0 & 0 & 0 & \Lambda & \Lambda & 0 \\ u & 0 & 0 & 0 & 0 & \Lambda & \Lambda & 0 \\ h & 0 & 0 & 0 & 0 & 0 & 0 & 0 \end{array} \begin{pmatrix} p_i \\ p_j \\ p_d \\ p_s \\ p_g \\ p_u \\ 0 \end{pmatrix} = \begin{pmatrix} \min(p_i, p_j) \\ \min(p_i, p_j) \\ p_d \\ p_s \\ \min(p_g, p_u) \\ \min(p_g, p_u) \\ 0 \end{pmatrix}.$$

Below, we illustrate the dynamics of the given $CAO_I(\mathbf{Q}^+)$ using a numerical example by parameterizing it and providing the initial data (Fig.6). Let $\#_i = 33i, \#_j = 21j, \forall \#_* = 0$.

Unlike *example 1*, here, with non-negative integer parameters of the cardinal semantic operators, the carries and the cardinals of CÆs are rational.

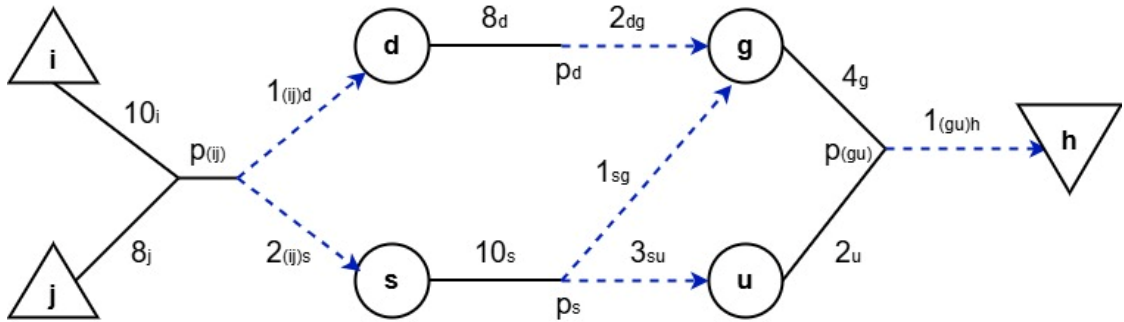

Fig. 6

This $CAO_I(\mathbf{Q}^+)$ is stationary. Its configuration matrix $\mathbf{\mathcal{P}}$ is expressed as

$$Ł = \begin{array}{c|ccccccc} \backslash & i & j & d & s & g & u & h \\ i & 10_i & 0 & 1_{(ij)d} & 2_{(ij)s} & 0 & 0 & 0 \\ j & 0 & 8_j & 1_{(ij)d} & 2_{(ij)s} & 0 & 0 & 0 \\ d & 0 & 0 & 8_d & 0 & 2_{dg} & 0 & 0 \\ s & 0 & 0 & 0 & 10_s & 1_{sg} & 3_{su} & 0 \\ g & 0 & 0 & 0 & 0 & 4_g & 0 & 1_{(gu)h} \\ u & 0 & 0 & 0 & 0 & 0 & 2_u & 1_{(gu)h} \\ h & 0 & 0 & 0 & 0 & 0 & 0 & 0 \end{array}.$$

Then, in accordance with the state equation (2), the state vector $|\#(k)\rangle$ at each step of the CST and the corresponding common carry vector $|p.(k)\rangle$ (control vector) are as follows:

$$
\begin{array}{cccc}
|\#(0)\rangle & |\#(1)\rangle & |\#(2)\rangle & |\#(3)\rangle \\
\begin{pmatrix} 33_i \\ 21_j \\ 0_d \\ 0_s \\ 0_g \\ 0_u \\ 0_h \end{pmatrix} \rightarrow &
\begin{pmatrix} 27/4_i \\ 0_j \\ 21/8_d \\ 21/4_s \\ 0_g \\ 0_u \\ 0_h \end{pmatrix} \rightarrow &
\begin{pmatrix} 27/4_i \\ 0_j \\ 0_d \\ 0_s \\ 189/160_g \\ 63/40_u \\ 0_h \end{pmatrix} \rightarrow &
\begin{pmatrix} 27/4_i \\ 0_j \\ 0_d \\ 0_s \\ 0_g \\ 63/64_u \\ 189/640_h \end{pmatrix}
\end{array}
$$

$$
\begin{array}{ccc}
|p.(0)\rangle & |p.(1)\rangle & |p.(2)\rangle \\
\begin{pmatrix} 21/8_{(ij)} \\ 21/8_{(ij)} \\ 0_d \\ 0_s \\ 0_{(gu)} \\ 0_{(gu)} \\ 0 \end{pmatrix} &
\begin{pmatrix} 0_{(ij)} \\ 0_{(ij)} \\ 21/64_d \\ 21/40_s \\ 0_{(gu)} \\ 0_{(gu)} \\ 0 \end{pmatrix} &
\begin{pmatrix} 0_{(ij)} \\ 0_{(ij)} \\ 0_d \\ 0_s \\ 189/640_{(gu)} \\ 189/640_{(gu)} \\ 0 \end{pmatrix}
\end{array}
$$

The full dynamics of the given $CAO_I(\mathbf{Q}^+)$ is determined by no more than three CST steps.

The values of the state vector components determine only the *meaning* of the CAO as a multicomponent formation. The semantic interpretation of the transformations is also determined by the position of each CÆ in the CAO structure as well as by the type and parameters of the cardinal semantic operators and their semantic connectivity (topology).

## 5. A First Step Towards Z

It would be nice to expand the class of "natural" Semantic Numeration Systems **SNS(N₀)** to the class of "whole numbers" Semantic Numeration Systems **SNS(Z)**, and then to the class of "rational" Semantic Numeration Systems **SNS(Q)**. However, this task is non-trivial, at least due to the problem of defining the common carry $p_{(.)}$.

Let's consider the case where the carries in any cardinal semantic operator are non-negative. We'll assume that some CÆs can "accept" not only positive values of transformants $q$, but also negative values $q^-$. This allows us to propose a class of SNS with adjustable cardinal values – **SNS(q⁻)**. A basic constraint must be introduced for any SNS(q⁻) ∈ **SNS(q⁻)**.

*Proposition.* As a result of any cardinal semantic transformation in SNS(q⁻), the values of all cardinals in a CAO must be non-negative.

The only way to form SNS(q⁻) is to assign negative values to the required conversion coefficients $r(.)$. This automatically entails negative values for the corresponding transformants $q(.)$.

Leaving the formalization for more detailed work, we limit ourselves here to an illustrative example.

***Example 3.*** Let a fragment of SNS(q⁻) be given by the following CAO formed by two L-operators with the initial data indicated on them (Fig. 7).

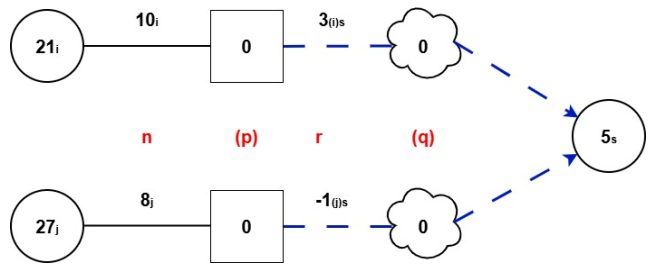

Fig.7

Let's find the values of the carries, remainders and transformants:

$p_i = \lfloor \#_i/n_i \rfloor = \lfloor 21_i/10_i \rfloor = 2_i, \quad p_j = \lfloor \#_j/n_j \rfloor = \lfloor 27_j/8_j \rfloor = 3_j;$
rem $\#_i = \#_i - p_i \cdot n_i = 21_i - 2_i \cdot 10_i = 1_i$, rem $\#_j = \#_j - p_j \cdot n_j = 27_j - 3_j \cdot 8_j = 3_j;$
$q_{(i)s} = r_{(i)s} \cdot p_i = 3_{(i)s} \cdot 2_i = 6_{(i)s}, \quad q_{(j)s} = r_{(j)s} \cdot p_j = -1_{(j)s} \cdot 3_j = -3_{(j)s};$

The intermediate results of the CST are presented in Fig.8-9.

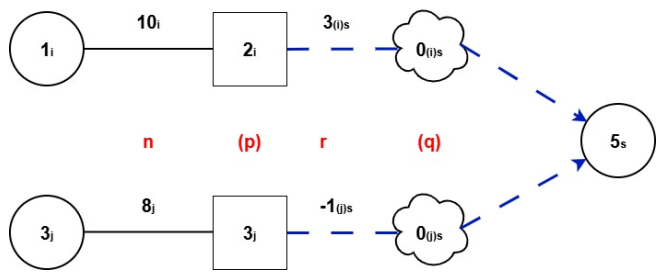

Fig.8

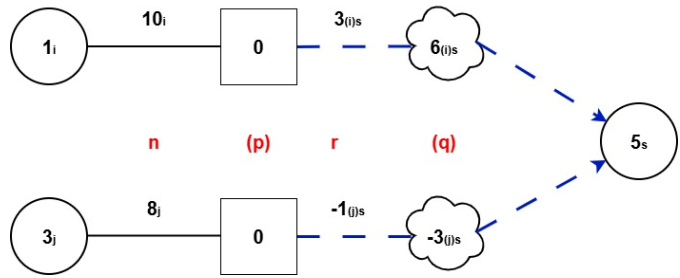

Fig.9

The outcome of the CST:

$$\text{rem } \#_i = 1_i, \text{ rem } \#_j = 3_j, \#'_s = \#_s + q_{(i)s} + q_{(j)s} = 5_s + 6_{(i)s} - 3_{(j)s} = 8_s$$

is presented in Fig.10.

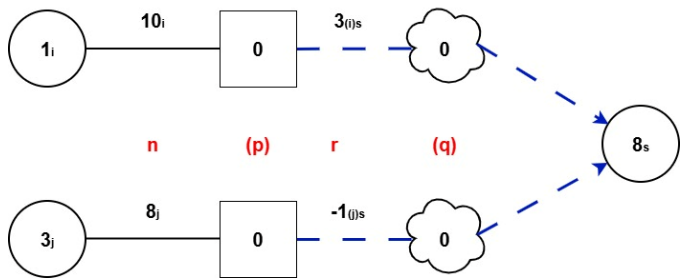

Fig.10

This approach allows us to describe controlled and adaptable dynamical systems in the Semantic Numeration Systems paradigm.

## 6. Conclusion

Any extension of a formal system typically raises more questions than it answers. The structural and parametric features of Semantic Numeration Systems make this approach not only complex but also fascinating. Furthermore, Semantic Numeration Systems are not simply an extensive expansion of traditional positional numeration systems, but distinctly diverse supersystems in all their structural and parametric diversity.

Rational Semantic Numeration Systems enable a more adequate modeling/description of real processes and transformations – not only in "whole portions" as in **SNS(N₀)**, but also in "pieces" and parts. However, for the rational representation to be complete, **SNS(Q⁻)** and, correspondingly, **SNS(Z)** are missing.

Are **SNS(Q)** and **SNS(Z)** fundamentally possible? For traditional numeration systems, represented only by L-operators in a linear connection topology, this is more of a computational question. For Semantic Numeration Systems, with their diverse topology and valence cardinal semantic operators, this question is non-trivial…